\documentclass[twocolumn,english,superscriptaddress,pra,floatfix,aps,10pt,nofootinbib]{revtex4-2}

\usepackage[colorlinks=true, allcolors=blue]{hyperref}
\usepackage[colorinlistoftodos, color=green!40, prependcaption]{todonotes}
\usepackage{comment}
\usepackage{geometry}
\usepackage{float}

\usetikzlibrary{shapes.geometric, arrows.meta, positioning, fit, calc}

\usepackage{algcompatible}
\usepackage{algpseudocode}   

\usepackage{tikz}
\usetikzlibrary{quantikz}
\usepackage{amsmath}

\usepackage{graphicx}
\usepackage{dcolumn}
\usepackage{bm}

\usepackage{newfloat}
\newgeometry{left=2cm, right=2cm, top=2cm, bottom=2cm} 

\DeclareFloatingEnvironment[
    fileext=loa,
    listname=List of Algorithms,
    name=ALGORITHM,
    placement=tbhp,
]{algorithm}

\begin{document}

\preprint{APS/123-QED}

\title{Methods for non-variational heuristic quantum optimisation}

\author{Stuart Ferguson}
 \email{S.A.ferguson-3@ed.ac.uk}
\author{Petros Wallden}%
 
\affiliation{%
 Quantum Software Lab, School of Informatics, The University of Edinburgh, Edinburgh, United Kingdom
}%

\date{\today}

\begin{abstract}

Optimisation plays a central role in a wide range of scientific and industrial applications, and quantum computing has been widely proposed as a means to achieve computational advantages in this domain.
To date, research into the design of noise-resilient quantum algorithms has been dominated by variational approaches, while alternatives remain relatively unexplored. In this work, we introduce a novel class of quantum optimisation heuristics that forgo this variational framework in favour of a hybrid quantum-classical approach built upon Markov Chain Monte Carlo (MCMC) techniques. We introduce Quantum-enhanced Simulated Annealing (QeSA) and Quantum-enhanced Parallel Tempering (QePT), before validating these heuristics on hard Sherrington-Kirkpatrick instances and demonstrate their superior scaling over classical benchmarks. These algorithms are expected to exhibit inherent robustness to noise and support parallel execution across both quantum and classical resources with only classical communication required. As such, they offer a scalable and potentially competitive route toward solving large-scale optimisation problems with near-term quantum devices.

\end{abstract}

\maketitle

\section{Introduction}

Early breakthroughs in quantum algorithms provided the rigorous proof-of-concept necessary to give birth to the field and motivate the development of quantum computing hardware \cite{grover1996, shor1994algorithms}.
Until recently, the only mechanism to research such algorithms was through their asymptotic computational complexity. However, as hardware advances, it is becoming increasingly important to investigate new heuristic methods for which, even classically, analytical guarantees are often lacking.

For the well-studied case of combinatorial optimisation, there is an abundance of classical solvers, both heuristic \cite{mart2018, juan2023review} and exact \cite{morrison2016branch}. In many cases, exact methods require exponential time in the worst case, meaning heuristic methods are relied upon to find sub-optimal solutions in a practical time-frame.

This classical analogy raises a question in quantum optimisation that is currently under-explored: What quantum heuristics exist \cite{buhrman2025formal}? Both Variational Quantum Algorithms (VQA) and Quantum Annealing (QA) are examples that have been extensively studied in the literature. However, in each case, the path forward may be unclear \cite{cerezo2021variational, larocca2025barren, rajak2023quantum, yarkoni2022quantum}. Some approaches to Quantum Computing Imaginary Time Evolution are also heuristics, but exhibit practical limitations in success probability or circuit depth \cite{motta2020determining, mcardle2019variational, kolotouros2025accelerating}. Recently there have been some other promising approaches that break this mould \cite{bennett2024, nakano2025fair, zhu2025combinatorial}. There are also some notable works that should be explicitly differentiated from our own: Quantum-enhanced optimisation (by warm starts) \cite{vcepaite2025quantum}; Quantum Simulated Annealing (fault tolerant) \cite{somma2007}; Quantum Replica Exchange \cite{chen2025quantum}; and Quantum-inspired Tempering \cite{albash2023quantum}.

The purpose of this paper is to propose a new class of quantum optimisation algorithms that take classical heuristics and apply a simple quantum subroutine. In particular, we discuss the use of real-time evolution which has previously supported the (up to) quartic quantum scaling advantage of Quantum-enhanced Markov Chain Monte-Carlo (QeMCMC) \cite{layden2023, christmann2025quantum, arai2025quantum, orfi2024bounding}. This is a very natural starting point as, for example, both Simulated Annealing (SA) and Parallel Tempering (PT) are algorithms that are fundamentally related to the MCMC. Here, we empirically compare these algorithms for the problem of combinatorial optimisation. Note that although PT is generally considered to be a sampling algorithm, it is often used for optimisation due to the link between sampling from complex ``low-temperature'' distributions and optimisation.

Our main contributions are the proposal and proof of principle of two novel quantum algorithms:

\begin{itemize} 
    \item Quantum-enhanced Simulated Annealing (QeSA) 
    \item Quantum-enhanced Parallel Tempering (QePT)
\end{itemize}

We also discuss some adaptions to these baseline approaches, incorporating ideas from classical methods, and introducing flexibility to the algorithms allowing for different hardware requirements and problem types. This motivates the necessity of High Performance Computing (HPC) hybridized with Quantum Computing (HPC-QC) for near-term quantum algorithms. Although we introduce these techniques, it is clear that they will not be well understood until they are tested on real quantum hardware. It is for this reason that we lay-out the algorithms, with some proof of principle numerics, but make no attempt to quantify any potential quantum speedup that may be possible with these methods.

The remainder of the paper is structured as follows. In Sec. \ref{sec:prelims} we revise combinatorial optimisation, MCMC, classical heuristics, QeMCMC and define the evaluation metrics we use. In Sec. \ref{sec:QEHO} we introduce our two new heuristics and provide some simulated numerics to motivate their future research. Finally, Sec. \ref{sec:discussion} discusses their potential and outlines the challenges surrounding their near-term implementation.

\section{Preliminaries}\label{sec:prelims}

In this section, we cover the background required for the rest of the paper. This includes a brief introduction to combinatorial optimisation and some detail on relevant problem instances, before relating optimisation to sampling. This link is the motivation for the algorithms described in this paper. We then describe classical sampling with MCMC, and the related classical optimisation heuristics, before introducing the quantum-enhanced approach to sampling, QeMCMC \footnote{In this paper, we have kept notation in line with \cite{layden2023} where possible.}. Finally, we highlight some numerical tools that can be used to assess the above algorithms, such ``spectral gap'' and computational ``effort''.

\subsection{Combinatorial optimisation}

In combinatorial optimisation, the aim is to find an optimal configuration ($s^*$) from all possible candidates, $s = [s_1,s_2,...,s_{n}]$, where each of the $n$ variables can take values $s_i \in \{+1, -1\}$. The set of candidates is the configuration space, $S$, with $|S| = 2^n$. Let the optimal solution be the configuration,

\begin{equation}
s^* = \text{argmin}_{s \in S}(f(s)),
\end{equation}
that minimises some objective function, $f(s)$. This formulation of the problem is equivalent to those which use binary instead of spin variables. Although our methods generalize to higher order objective functions, focus is normally placed on Quadratic Unconstrained Binary Optimisation (QUBO) problems in the context of near-term quantum optimisation.

We consider one class of functions called Sherrington-Kirkpatrick models, which are fully-connected, second-order spin glasses with objective function

\begin{equation}
\label{eqn:comb_opt}
f_{\text{SK}}(s) = -\sum_{i=1}^n J^{(1)}_i s_i - \sum_{i<j}^n J^{(2)}_{ij} s_i s_j.
\end{equation}

Each element in $J^{(1)}$ and $J^{(2)}$ is drawn from a standard normal distribution. This corresponds to a particularly hard case of the general class of problems, as the cost function is ``glassy'' \textemdash meaning that it has many disparate local minima. In our proof of principle numerics we study $f_{SK}$ for $n \leq 10$, generating random model instances by sampling $J^{(1)}$ and $J^{(2)}$ from the normal distribution. We compare approaches to combinatorial optimisation by averaging the performance of each algorithm over at least $100$ such instances. The choice to use Sherrington-Kirkpatrick models is motivated by the glassiness inherent in most problem instances, resulting in small, yet difficult, computational problems that showcase the potential of our techniques. Although we leave analysis of other problem types to future work, it should be noted that Ising models such as this are very general and sub-cases include canonical tasks such as Maxcut and the Travelling Salesperson Problem.

\subsection{MCMC}
The spin formulation of our problem is motivated by the canonical description of the main underlying computational technique: Markov Chain Monte Carlo. MCMC (Alg. \ref{alg:MCMC}) is designed to approximately \textit{sample} from probability distributions that are intractable \cite{hastings1970, metropolis1953}.

\begin{algorithm}
\caption{MCMC}
\begin{algorithmic}[0] 
    \State 
    \State $s \gets \text{initial spin configuration}$
    
    \While{not converged}
        \State $s^{\prime} \gets \text{LocalProposal}(s)$
        \State $A(s^{\prime} \mid s) \gets \min \left(1, \frac{\mu(s^{\prime})}{\mu(s)} \frac{Q(s \mid s^{\prime})}{Q(s^{\prime} \mid s)}\right)$
        \State $u \gets \text{Uniform}(0,1)$
        \If{$u < A(s^{\prime} \mid s)$}
            \State $s \gets s^{\prime}$
        \EndIf
    \EndWhile
\end{algorithmic}
\label{alg:MCMC}
\end{algorithm}

One example is the Boltzmann distribution,

\begin{equation}
    \mu(s)=\frac{1}{Z} e^{-f(s) / T},
    \label{eqn:mu}
\end{equation}

where $Z$ is the partition function, $Z = \sum_{s\in S} e^{-f(s) / T}$ and $T$ is a temperature parameter. Exact calculation of $Z$ is intractable - requiring $\Omega\left(2^n\right)$ time. MCMC makes no attempt to approximate $Z$, and instead performs a random walk on $S$. 
The probability of any transition $s \rightarrow s^\prime$, $P(s^{\prime} | s)$, must be chosen so that after many steps, the chance of finding the chain in any state, $s$, is approximately $\mu(s)$. To ensure convergence to this stationary distribution the chain must be both irreducible and aperiodic, while satisfying a condition known as detailed balance,

\begin{equation}
    P(s^{\prime} \mid s) \mu(s) = P(s^{\prime} \mid s) \mu(s^\prime).
\end{equation}

The usual approach to ensure detailed balance is to take two distinct actions within a transition: first \emph{propose} a transition from the current  ($s$) to a new ($s^\prime$) configuration with probability $Q(s^\prime|s)$; then \emph{accept} the new proposal with probability $A(s^\prime|s)$. The resulting probability, 

\begin{equation}
    P(s^{\prime} \mid s)=A(s^{\prime} \mid s) Q(s^{\prime} \mid s),
\end{equation}

can satisfy detailed balance for a variety of acceptance criteria, but Metropolis Hastings, 

\begin{equation}
\label{eqn:MH}
A(s^{\prime} \mid s)=\min (1, \frac{\mu(s^{\prime})}{\mu(s)} \frac{Q(s \mid s^{\prime})}{Q(s^{\prime} \mid s)})= \min (1,e^{-\frac{\Delta f}{T}}),
\end{equation}

is the most common. The second equality comes from the cancellation of partition functions and the fact that common proposal methods, such as simply flipping a single spin (local proposal, Alg. \ref{alg:local_proposal}), are chosen to ensure $Q(s^{\prime} \mid s) =  Q(s \mid s^{\prime})$. If one is able to approximately sample from the Boltzmann distribution, then by setting $T \approx  0$ they can solve the related optimisation problem, Eq. \ref{eqn:comb_opt}. Naturally, however, the computational cost of MCMC is extremely temperature dependent, and in the low-temperature regime it is empirically understood that many steps are required to converge to the stationary distribution: \emph{thermalisation} is extremely slow.

\begin{algorithm}
\caption{Local Proposal}
\label{alg:local_proposal}
\begin{algorithmic}[1]
\Function{LocalProposal}{$s$}
\State $s^{\prime} \gets \text{copy}(s)$ \Comment{Avoid in place modification}
\State $n \gets \text{length}(s)$
\State $i \gets \text{random integer in } [1, n]$ \Comment{Select a random spin}
\State $s^{\prime}_i \gets -s^{\prime}_i$ \Comment{Flip the spin at the chosen site}
\State \Return $s^{\prime}$
\EndFunction
\end{algorithmic}
\label{alg:proposal}
\end{algorithm}

\subsection{Classical optimisation heuristics}

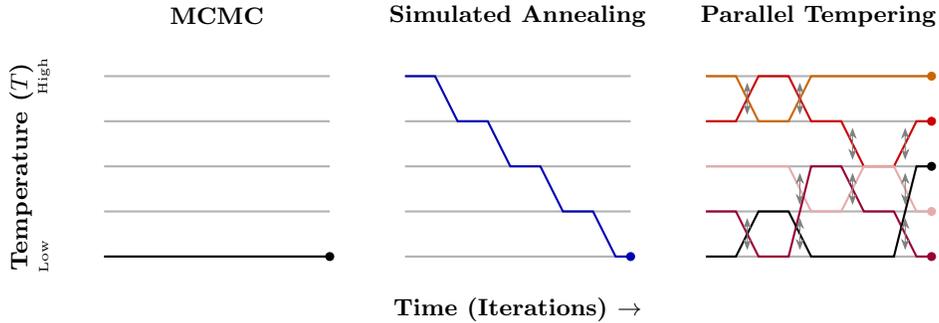
\begin{figure*}[!t]
    \centering
    \colorlet{T0}{black}               
\colorlet{T1}{purple!80!black} 
\colorlet{T2}{pink!90!black}
\colorlet{T3}{red!80!black}
\colorlet{T4}{orange!80!black}    
\begin{tikzpicture}
        \tikzset{
            rung/.style={thick, gray!60},
            activeRung/.style={thick, gray!50},
            particle/.style={circle, fill=#1, inner sep=1.2pt},
            pathline/.style={thick, #1}
        }

        \node[rotate=90, font=\small\bfseries] at (-1.1, 1.2) {Temperature ($T$)};
        \foreach \y/\label in {0/Low, 2.4/High} \node[rotate=90, font=\tiny] at (-0.8, \y) {\label};

        \begin{scope}[shift={(0,0)}]
            \node at (1.5, 3.2) {\textbf{MCMC}};
            \foreach \y in {0, 0.6, 1.2, 1.8, 2.4} \draw[rung] (0,\y) -- (3,\y);
            \draw[activeRung] (0,0) -- (3,0); 
            
            \draw[pathline=T0] (0,0) -- (3,0);

            \node[particle=T0] at (3,0) {};
        \end{scope}

        \begin{scope}[shift={(4,0)}]
            \node at (1.5, 3.2) {\textbf{Simulated Annealing}};
            \foreach \y in {0, 0.6, 1.2, 1.8, 2.4} \draw[rung] (0,\y) -- (3,\y);
            
            \draw[pathline=blue!70!black] (0,2.4) -- (0.4,2.4) -- (0.7,1.8) -- (1.1,1.8) -- (1.4,1.2) -- (1.8,1.2) -- (2.1,0.6) -- (2.5,0.6) -- (2.8,0) -- (3,0);
            
            \node[particle=blue!70!black] at (3,0) {};
        \end{scope}

        \begin{scope}[shift={(8,0)}]
            \node at (1.5, 3.2) {\textbf{Parallel Tempering}};
            \foreach \y in {0, 0.6, 1.2, 1.8, 2.4} \draw[rung] (0,\y) -- (3,\y);

            \draw[pathline=T4] (0,2.4) -- (0.4,2.4) -- (0.7,1.8) -- (1.1,1.8) -- (1.4,2.4) -- (3,2.4);
            \draw[pathline=T3] (0,1.8) -- (0.4,1.8) -- (0.7,2.4) -- (1.1,2.4) -- (1.4,1.8) -- (1.8,1.8) -- (2.1,1.2) -- (2.5,1.2) -- (2.8,1.8) -- (3,1.8);
            \draw[pathline=T2] (0,1.2) -- (1.1,1.2) -- (1.4,0.6) -- (1.8,0.6) -- (2.1,1.2) -- (2.5,1.2) -- (2.8,0.6) -- (3,0.6);
            \draw[pathline=T1] (0,0.6) -- (0.4,0.6) -- (0.7,0) -- (1.1,0) -- (1.4,1.2) -- (1.8,1.2) -- (2.1,0.6) -- (2.5,0.6) -- (2.8,0) -- (3,0);
            \draw[pathline=T0] (0,0) -- (0.4,0) -- (0.7,0.6) -- (1.1,0.6) -- (1.4,0) -- (2.5,0) -- (2.8,1.2) -- (3,1.2);

            \draw[<->, >=Stealth, ultra thin, gray] (0.55, 1.9) -- (0.55, 2.3); 
            \draw[<->, >=Stealth, ultra thin, gray] (1.25, 1.9) -- (1.25, 2.3); 
            
            \draw[<->, >=Stealth, ultra thin, gray] (1.95, 1.3) -- (1.95, 1.7); 
            \draw[<->, >=Stealth, ultra thin, gray] (2.65, 1.3) -- (2.65, 1.7); 

            \draw[<->, >=Stealth, ultra thin, gray] (1.25, 0.7) -- (1.25, 1.1); 
            \draw[<->, >=Stealth, ultra thin, gray] (1.95, 0.7) -- (1.95, 1.1); 
            \draw[<->, >=Stealth, ultra thin, gray] (2.65, 0.7) -- (2.65, 1.1); 
            
            \draw[<->, >=Stealth, ultra thin, gray] (0.55, 0.1) -- (0.55, 0.5); 
            \draw[<->, >=Stealth, ultra thin, gray] (1.25, 0.1) -- (1.25, 0.5); 
            \draw[<->, >=Stealth, ultra thin, gray] (2.65, 0.1) -- (2.65, 0.5);

            \foreach \y/\c in {0/T1, 0.6/T2, 1.2/T0, 1.8/T3, 2.4/T4} \node[particle=\c] at (3,\y) {};
        \end{scope}

        \node[font=\small\bfseries] at (5.5, -0.7) {Time (Iterations) $\to$};

    \end{tikzpicture}
    \caption{Visual representation of MCMC, SA and PT. MCMC attempts to thermalize a single Markov chain of a given temperature, $T_{\text{low}}$. SA starts at a high temperature and anneals to $T_{\text{low}}$, while PT runs multiple chains in parallel at different temperatures, swapping configurations between chains occasionally. }
    \label{fig:sampling_comparison}
\end{figure*}

One naive approach to optimisation would be to run a low-temperature MCMC. It is well known that in spin-glass Ising models, as $T \rightarrow 0$, the probability of sampling the ground state is maximized. However, the number of steps required to thermalize low-temperature chains scales exponentially in $n$, and thus becomes completely intractable even for only relatively small systems \cite{levin2017}. This is primarily due to the ``energy barriers'' in the cost function that separate local minima. At low temperatures, the probability of accepting a proposed transition that increases the energy becomes vanishingly small.

To mitigate the issue of being trapped in local minima, SA (Alg. \ref{alg:sa}) introduces a schedule ($T_\text{schedule}$), which incrementally reduces $T$ at each step of the algorithm \cite{bertsimas1993simulated, kirkpatrick1983optimization}. If the initial temperature ($T_0$) is high enough to allow the chain to traverse the cost function landscape globally, and the schedule is slow enough, then the chain is highly likely to find the minima. In this paper, we assume a logarithmic cooling schedule.

\begin{algorithm}
\caption{Simulated Annealing }
\begin{algorithmic}[0]

  \State $s \gets$ Initial spin configuration
  \State $T_\text{schedule} \gets $ Temperature Schedule
  
  \For{$T$ in $T_{schedule}$}
      \State $s' \gets$ LocalProposal(s)
      \State $\Delta f \gets f(s') - f(s)$
      \State $A(s^{\prime} \mid s) \gets \min \left(1, \frac{\mu(s^{\prime})}{\mu(s)} \frac{Q(s \mid s^{\prime})}{Q(s^{\prime} \mid s)}\right)$

      \State $u \gets \text{Uniform}(0,1)$
      \If{$u < A(s^\prime \mid s)$}
        \State $s \gets s'$
      \EndIf
    \EndFor
  \State \Return $s$
\end{algorithmic}
\label{alg:sa}
\end{algorithm}

It should be stressed that the sole difference between MCMC and SA, is that the temperature is no longer fixed. This extremely simple change transforms the process into a powerful optimisation algorithm. However, due to the necessarily serial nature of Markov chains, highly parallelisable modern approaches typically outperform SA.

One method that neatly parallelizes MCMC methods is PT, Alg. \ref{alg:pt} and is currently used in HPC for both sampling and optimisation. Rather than using just a single Markov chain, $M$ ``replicas'' are run in parallel, each at fixed temperature, $T_1 < T_2 < ... < T_M$. Periodically an attempt is made to exchange their configurations - hence the alias ``replica-exchange MCMC''.

\begin{algorithm}
\caption{Parallel Tempering}
\label{alg:pt}\begin{algorithmic}[0]
\State $T_{ladder} \gets \text{Temperature ladder}$
\State $\{s^{(1)}, s^{(2)}, \dots, s^{(M)}\} \gets$ $M$ independent initial spin configurations
\State $k \gets$ Swap interval
\State $\text{step} \gets 0$ \Comment{Timesteps}
\While{not converged}
\State $\text{step} \gets \text{step} + 1$\For{$i = 1$ \textbf{to} $M$} \Comment{Parallel MCMC Step}

\State $s^{(i)^\prime} \gets \text{LocalProposal}(s^{(i)})$\State $\Delta f \gets f(s_i') - f(s_i)$
\State $A(s^{(i)^\prime} \mid s^{(i)}) \gets \min \left(1, \frac{\mu(s^{\prime})}{\mu(s)} \frac{Q(s^{(i)} \mid s^{(i)^\prime}}{Q(s^{(i)^\prime} \mid s^{(i)})}\right)$
\State $u \gets \text{Uniform}(0,1)$
\If{$u < A(s^{(i)^\prime} \mid s^{(i)})$}

\State $s^{(i)} \gets s^{(i)^\prime}$
\EndIf
\EndFor
\If{$\text{step} \pmod k == 0$} \Comment{Replica Exchange Step}
    \State $start \gets 2 - ((\text{step}/k) \pmod 2)$ 
    \For{$i = start, start+2, \dots$ while $i < M$}
        \State $j \gets i + 1$
        \State $\Delta \beta \gets (1/T_i - 1/T_j)$
        \State $\Delta f_{ij} \gets f(s^{(i)}) - f(s^{(j)})$
        \State $v \gets \text{Uniform}(0,1)$
        \If{$v < \exp(\Delta \beta \cdot \Delta f_{ij})$}
            \State \text{Swap } $s^{(i)} \leftrightarrow s^{(j)}$
        \EndIf
    \EndFor
\EndIf
\EndWhile
\State \Return $s$ \text{associated with } $T_{\min}$
\end{algorithmic}
\end{algorithm}

While various exchange schemes exist, we consider a common approach where exchanges are attempted between adjacent replicas at consistent intervals of $k$ steps. To ensure all pairs have the opportunity to swap, the algorithm alternates between two sets of pairs. At the first interval, swaps are attempted for all ``even'' pairs $(s^{(i)}, s^{({i+1})})$ where $i \in \{0, 2, 4, \dots\}$ and $s^{(i)}$ is the configuration of the i\textsuperscript{th} replica. At the next interval, swaps are attempted for ``odd'' pairs where $i \in \{1, 3, 5, \dots\}$. A proposed swap between replica $i$ (at temperature $T_i$) and replica $j$ is accepted with probability:

\begin{equation} 
A^{\text{PT}}(s^{(i)}, s^{(j)}) = \min\Big[1, \exp\left((\frac{1}{T_i} - \frac{1}{T_j})(f(s^{(i)}) - f(s^{(j)}))\right)\Big].
\end{equation}

In this way, any configuration can migrate ergodically between replicas. In analogy with $T_\text{schedule}$ in SA, PT requires the user to choose the temperature of each replica, $T_{ladder}$. For consistency we will again choose this to be logarithmic. PT also requires a choice of number of replicas, which if chosen poorly, can lead to inefficient implementation. If $M$ is too large then inter-chain mixing is slow, while in the opposite extreme the energy distributions of adjacent replicas will have negligible overlap.

\subsection{QeMCMC}

Low-temperature sampling is one of the core motivations for the quantum-enhanced approach introduced in \cite{layden2023}. By describing a unitary that satisfies $|\langle s^{\prime}| U| s\rangle|=|\langle s| U| s^{\prime}\rangle|$, a Quantum-enhanced MCMC \emph{proposal} mechanism can be built by simply initialising a quantum computer in a computational basis state\footnote{This is done by mapping the spin variables to binary variables.} $|s\rangle$, applying $U$, and measuring the system in the computational basis. 

However, the question is, can a simple unitary be designed that is helpful? Heuristically, the optimal proposal has two distinct properties: small $\Delta f$; and large Hamming distance\footnote{We define Hamming distance as the difference in Hamming weight between $s$ and $s^\prime$.} ($\Delta H$). Small $\Delta f$ means $A(s^\prime|s)\approx1$ so the chain moves consistently, and large $\Delta H$ means that a chain moves non-locally between configurations that are structurally very different. One approach introduced by \cite{layden2023} that has these properties is real time evolution $U = \exp^{-itH}$ under a Hamiltonian,

\begin{equation}
H = \gamma \big[\sum_{j=1}^n X_j\big] -(1- \gamma)\alpha \big[\sum_{j>k=1}^n J^{(1)}_{j k} Z_j Z_k +\sum_{j=1}^n J^{(2)}_j Z_j\big] ,
\label{eqn:H_tot}
\end{equation}

where $Z$ is the Pauli matrix, and $\gamma$ and $t$ are hyper-parameters that should be carefully selected. The second term in this Hamiltonian encodes the Ising model instance, while the first is related to the driving term in adiabatic quantum computing: it forces the system to spread among nearby energy levels. This means that $\gamma$ can be tuned to produce a resulting quantum state that has large amplitude on basis states for which $\Delta f$ is small. When this quantum state is measured, the resulting configuration, $s^\prime$, has the property $f(s)\approx f(s^\prime)\pm \alpha$ for small $\alpha$. The proposal function is detailed in Alg. \ref{alg:quantum_proposal}.

\begin{algorithm}
\caption{Quantum Proposal}
\label{alg:quantum_proposal}
\begin{algorithmic}[1]
\Function{QuantumProposal}{$s$}
    \State $t \gets \text{Uniform}(t_{\min}, t_{\max})$ \Comment{Evolution time}
    \State $\gamma \gets  \text{Uniform}(\gamma_{\min}, \gamma_{\max})$ \Comment{Mixing}
    \State \textbf{Apply Circuit:}
    \State 
    \begin{center}
        \begin{tikzcd}
            \lstick{$\ket{s}$} & \qwbundle{n} & \gate{e^{-iH(\gamma)t}} & \meter{} & \cw \rstick{$s'$}
        \end{tikzcd}
    \end{center}
    \State \Return $s^{\prime}$
\EndFunction
\end{algorithmic}
\end{algorithm}

This heuristic design means that performance is not well understood analytically. The thermalization time ($\tau$) of a given MCMC is clearly dependent on the starting state, and in practise it is not always clear from sampling a chain whether it is in the stationary state. However, empirical results suggest that the quantum scaling advantage can be somewhere between cubic and quartic depending on the temperature setting.

\subsection{Evaluation metrics}

An important quantity is the spectral gap, $\delta=1-\max _{\lambda \neq 1}|\lambda|$, where $\{\lambda\}$ are the eigenvalues of the $2^n \times 2^n$ matrix $P$. The spectral gap bounds $\tau$ with respect to $\varepsilon$, the total variational distance between the distribution of the Markov chain and $\mu(s)$ \cite{levin2017}.

\begin{equation}
\left(\frac{1}{\delta}-1\right) \ln \left(\frac{1}{2 \varepsilon}\right) \leq \tau \leq \frac{1}{\delta} \ln \left(\frac{1}{\varepsilon \min _s \mu(s)}\right).
\label{eqn:spec_therm}
\end{equation}

Thus, by understanding the exponential decay $\delta$ with $n$, the scaling of $\tau$ can be understood for a given proposal without ever running an MCMC. Of course, calculating $\delta$ is trivially intractable, so this method can only be used for small systems. The spectral gap can also be related directly to optimisation in some cases - see Sec. \ref{sec:QeSA_pop}.

For the case of evaluating the success of optimisation algorithms, we choose to define the computational effort ($N_{p}$) that has gone into finding the global minima with probability $p$. In our case, the universal unit for ``effort'' is a single use of the proposal mechanism - or equivalently an evaluation of $f$. Of course, a quantum strategy will require a considerable amount of time to perform a proposal, however this is hardware specific. The total effort for given approach is $N_{p} =\ell \times R$, the length of a single Markov chain ($\ell$), multiplied by the number ($R$) of repeats required to find the ground state with probability $p=1-\epsilon$ \cite{ronnow2014,herr2017optimizing}:

\begin{equation}
\label{eqn:effort}
R=\log (1-p) / \log (1-p_s).
\end{equation}

Where $p_s$ is the success probability of a single run of the optimisation algorithm which can be approximated numerically for small $n$. If we have multiple replicas, for example in PT, then the effort of a single run is the product of number of chains and effort of a single chain $\ell_{PT} = M\ell$. Note that this is a slight underestimate, as replica exchanges have an associated cost but the effect is negligible. The rest of the paper will refer to the standard $\epsilon = 0.01$, $N_{0.99}$.

This definition of effort means that we can fairly compare different approaches to optimisation. For medium scale problems, we can approximately optimise hyper-parameters to minimize the effort, meaning that methods can then be fairly compared, in terms of their ``optimal'' effort.

\section{Quantum-enhanced optimisation} \label{sec:QEHO}
In this section, we introduce the two novel quantum algorithms, Quantum-enhanced Parallel Tempering, and Quantum-enhanced Simulated Annealing. We first motivate the existence of these quantum heuristics in particular against related fault-tolerant quantum algorithms. We then provide proof of principle numerics that show their favourable convergence compared to classical approaches.

\subsection{QeSA}\label{sec:QeSA}

In the same way that SA is an incredibly effective way to extend MCMC to optimisation, QeSA extends QeMCMC. The idea is simple: run QeMCMC while annealing the temperature parameter. Done carefully, the probability of descending to the local minima should be higher than that of a comparable SA. Naturally, the Quantum-enhanced convergence can be expected to increase mobility at the late (low-temperature) stage of annealing. See Alg. \ref{alg:QeSA} for details.

\begin{algorithm}
\caption{Quantum-enhanced Simulated Annealing }
\begin{algorithmic}[0]

  \State $s \gets$ Initial spin configuration
  \State $T_\text{schedule} \gets $ Temperature Schedule

  \For{$T$ in $T_{schedule}$}
      \State $s' \gets$ QuantumProposal(s)
      \State $\Delta f \gets f(s') - f(s)$
      \State $A(s^{\prime} \mid s) \gets \min \left(1, \frac{\mu(s^{\prime})}{\mu(s)} \frac{Q(s \mid s^{\prime})}{Q(s^{\prime} \mid s)}\right)$

      \State $u \gets \text{Uniform}(0,1)$
      \If{$u < A(s^\prime \mid s)$}
        \State $s \gets s'$
      \EndIf
    \EndFor
  \State \Return $s$
\end{algorithmic}
\label{alg:QeSA}
\end{algorithm}

It is important to highlight the clear differences between our approach and the quadratic quantum speed-up of the fault tolerant Quantum Simulated Annealing (QSA) algorithm \cite{somma2007}. Our method is hybrid, with the Markov chain being stored primarily in classical compute and only the proposal process being quantum. The previous approach, however, uses fault tolerant techniques, including quantum phase estimation, to repeatedly project onto the quantum Gibbs state for the annealing temperature.
In the process of deriving their quantum speed-up, the runtime ($N^{SA}$, number of total proposals) of any classical SA algorithm is upper bounded,

\begin{equation}
\label{eqn:SA_delta}
\mathcal{O}\left(\log \left(d / \epsilon_s^2\right) / \delta_{min}\right),
\end{equation}

where $d = 2^n$ is the dimension of the state space, $p_s = 1-\epsilon_s$ is the success probability of a single run, and $\delta_{min} = \text{min}_{T\in T_\text{schedule}}(\gamma)$ is the minimum spectral gap over all temperatures in the schedule. The equivalent for QSA is 

\begin{equation}
\label{eqn:QSA_delta}
\mathcal{O}\left(\log ^3\left(d / \epsilon_s^2\right) /\left(\epsilon_s^2\sqrt{\delta_{min}} \right)\right)
\end{equation}

resulting in a quadratic speed-up in $\delta_{\text{min}}$. By comparing Eqs. \ref{eqn:SA_delta} and \ref{eqn:QSA_delta}, it should be clear that the proposed QeSA derives quantum speed-up in an entirely different manner from its fault-tolerant ancestor. QeSA is still bound by the Eq. \ref{eqn:SA_delta}, but it directly reduces $\delta_\text{min}$ through the QeMCMC proposal. Note that the reduction in spectral gap of the QeMCMC is only for average case instances, and it has been proven ineffective in the worst case \cite{orfi2024bounding}.

Thus, the main motivation for QeSA is that the low-temperature convergence of QeMCMC could result in $N^{QeSA}<N^{SA}$. We thus hypothesise an \emph{up to quartic scaling advantage in combinatorial optimisation} for average case spin glass Ising models. 

    \subsubsection{QeSA Proof of principle} \label{sec:QeSA_pop}

To test this hypothesis, we perform experiments where QeSA is trialled in a simulated environment. Of course, one must choose an optimal annealing schedule to fairly compare SA and QeSA. Here, we choose $T_{\text{high}} = 10$ and $T_{\text{low}} = 0.1$, and annealing schedule of the form $T_i = T_{\text{high}} \times e^{-ax}$. We parameterise the length of an annealing run by the number of steps the Markov chain takes, $\ell$, which is of course inversely related to $a$. In all experiments, $100$ model instances are studied, with $100$ annealing runs for each data point. The quantum-enhanced hyper-parameters are sampled from the ranges: $\left[\gamma_{\min }, \gamma_{\max }\right]=[0.25,0.6]$, and $\left[t_{\min }, t_{\max }\right]=[2,20]$. Note that $t$ represents an integer number of Trotter time-steps, each of length $\Delta t = 0.8$ as in \cite{layden2023}. A simple implementation of the algorithm is included in \cite{Ferguson_Quantum-enhanced_Simulated_Annealing}.

\begin{figure}[h]
    \centering
    \includegraphics[width=1\linewidth]{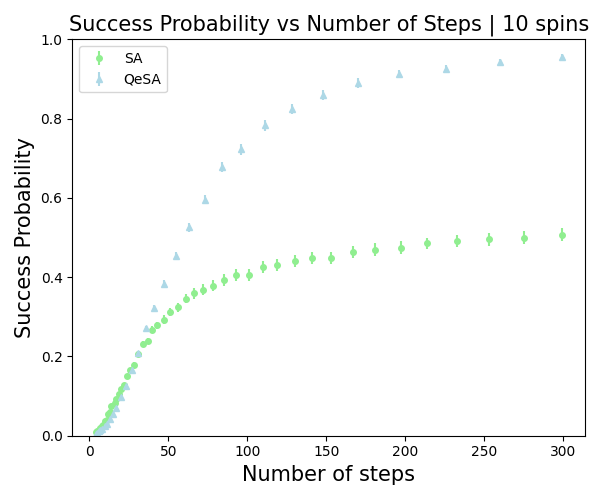}
    \caption{The probability of finding global minima, against the number of steps taken. Classical SA (green) quickly finds the global minima occasionally, however if it is not in the local region of the initial state, then it gets stuck in local minima and can take an extremely long time to escape. QeSA (blue), however, does not suffer from such locality, and longer annealing runs are rewarded with a very high probability of finding the global minima. Each data point is the average of 100 different models, where 100 anneals are performed for each step value. }
    \label{fig:10_hopsprobs}
\end{figure}

Of course, one must consider that there is a choice to be made between repeating many cheap anneals, or doing fewer longer annealing runs. As seen in Fig. \ref{fig:10_hopsprobs}, the classical SA is occasionally able to quickly find the global minima, but often gets stuck in local minima. Conversely, the quantum approach does not suffer from such locality issues and consistently reaches much higher success probability after only a small increase in number of steps.

\begin{figure}[h]
    \centering
    \includegraphics[width=1\linewidth]{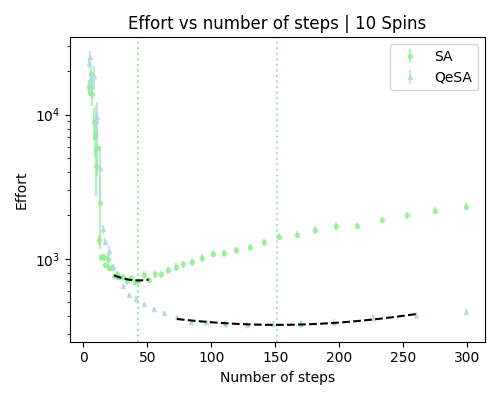}
    \caption{Effort (Eq. \ref{eqn:comb_opt}) required to find the global minima against the number of Markov chain steps for $n=10$ for SA (green) and QeSA (blue). Although the classical approach works well when repeating many short anneals, the quantum approach can quickly reduce the effort by performing fewer, longer anneals. A subset of low effort values are selected and a simple quadratic function (black) over these are used to approximate the optimal effort in each case. The optimal number of steps, represented by vertical dotted lines, are $\ell_{SA} \approx 43$ and $\ell_{QeSA} \approx 151$ }
    \label{fig:10_effort}
\end{figure}

\begin{figure}[h]
    \centering
    \includegraphics[width=1\linewidth]{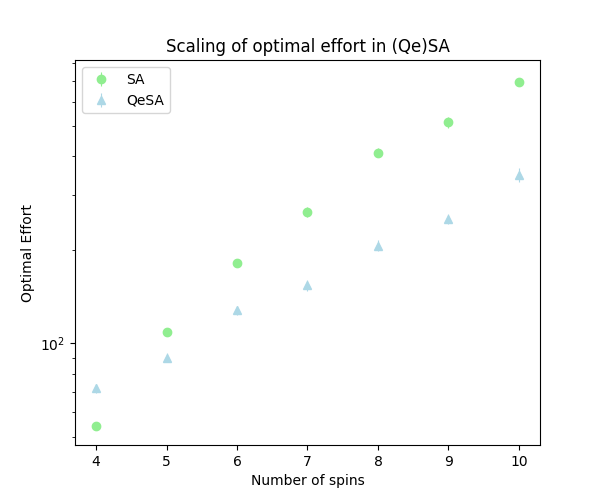}
    \caption{Optimal effort for SA (green, circle) and QeSA (blue, triangle) as the number of spins is increased. For each $n$, the optimal $\ell$ is found (as in Fig. \ref{fig:10_effort}), and $100$ new randomly initialised Ising models are then used to generate this figure, with each anneal now being repeated $1000$ times.}
    \label{fig:SA_effort}
\end{figure}

This means that we must optimise $\ell$ with respect to the total computational effort,  Eq. \ref{eqn:effort}, to fairly compare methods. An example of this is shown for $n = 10$ in Fig. \ref{fig:10_effort}.
Once good values of $\ell$ are found for each problem size, we can directly compare the required computational effort between QeSA and SA \textemdash Fig. \ref{fig:SA_effort}.

The form of this scaling relationship is not entirely clear from Fig. \ref{fig:SA_effort} for both SA and QeSA, so we are careful to not make any assumptions. It is clear, however, that our quantum approach has the potential to outperform its classical alternative.

\subsection{QePT} \label{sec:QePT}

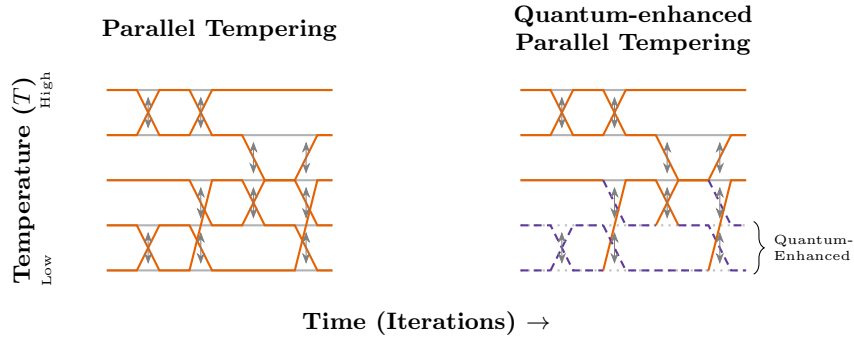
\begin{figure*}[t]
    \centering
    \begin{tikzpicture}
    \definecolor{C_Classical}{HTML}{E66100} 
    \definecolor{C_Quantum}{HTML}{5D3A9B}   

    \tikzset{
        rung/.style={thick, gray!60},
        activeRung/.style={thick, gray!50},
        quantumRung/.style={thick, gray!50, dotted},
        particle/.style={circle, fill=#1, inner sep=1.2pt},
        classicalPath/.style={thick, C_Classical},
        quantumPath/.style={thick, C_Quantum, dash pattern=on 3pt off 2pt}   }

    \node[rotate=90, font=\small\bfseries] at (-1.1, 1.2) {Temperature ($T$)};
    \foreach \y/\label in {0/Low, 2.4/High} \node[rotate=90, font=\tiny] at (-0.8, \y) {\label};

    \begin{scope}[shift={(0,0)}]
        \node at (1.5, 3.2) {\textbf{Parallel Tempering}};
        \foreach \y in {0, 0.6, 1.2, 1.8, 2.4} \draw[rung] (0,\y) -- (3,\y);

        \draw[classicalPath] (0,2.4) -- (0.4,2.4) -- (0.7,1.8) -- (1.1,1.8) -- (1.4,2.4) -- (3,2.4);
        \draw[classicalPath] (0,1.8) -- (0.4,1.8) -- (0.7,2.4) -- (1.1,2.4) -- (1.4,1.8) -- (1.8,1.8) -- (2.1,1.2) -- (2.5,1.2) -- (2.8,1.8) -- (3,1.8);
        \draw[classicalPath] (0,1.2) -- (1.1,1.2) -- (1.4,0.6) -- (1.8,0.6) -- (2.1,1.2) -- (2.5,1.2) -- (2.8,0.6) -- (3,0.6);
        \draw[classicalPath] (0,0.6) -- (0.4,0.6) -- (0.7,0) -- (1.1,0) -- (1.4,1.2) -- (1.8,1.2) -- (2.1,0.6) -- (2.5,0.6) -- (2.8,0) -- (3,0);
        \draw[classicalPath] (0,0) -- (0.4,0) -- (0.7,0.6) -- (1.1,0.6) -- (1.4,0) -- (2.5,0) -- (2.8,1.2) -- (3,1.2);

        \foreach \x/\y in {0.55/1.9, 1.25/1.9, 1.95/1.3, 2.65/1.3, 1.25/0.7, 1.95/0.7, 2.65/0.7, 0.55/0.1, 1.25/0.1, 2.65/0.1}
            \draw[<->, >=Stealth, ultra thin, gray] (\x, \y) -- (\x, \y+0.4);
    \end{scope}

    \begin{scope}[shift={(5.5,0)}]
\node[align=center] at (1.5, 3.2) {\textbf{Quantum-enhanced}\\\textbf{Parallel Tempering}};        
        \draw[quantumRung] (0,0) -- (3,0);
        \draw[quantumRung] (0,0.6) -- (3,0.6);
        \foreach \y in {1.2, 1.8, 2.4} \draw[rung] (0,\y) -- (3,\y);

        \draw [decorate, decoration={brace, amplitude=4pt, mirror}] (3.1,-0.05) -- (3.1,0.65) 
            node [black, midway, xshift=0.8cm, font=\tiny, align=left] {Quantum-\\Enhanced};

        \draw[classicalPath] (0,2.4) -- (0.4,2.4) -- (0.7,1.8) -- (1.1,1.8) -- (1.4,2.4) -- (3,2.4);
        
        \draw[classicalPath] (0,1.8) -- (0.4,1.8) -- (0.7,2.4) -- (1.1,2.4) -- (1.4,1.8) -- (1.8,1.8) -- (2.1,1.2) -- (2.5,1.2) -- (2.8,1.8) -- (3,1.8);
        
        \draw[classicalPath] (0,1.2) -- (1.1,1.2);
        \draw[quantumPath] (1.1,1.2) -- (1.4,0.6) -- (1.8,0.6);
        \draw[classicalPath] (1.8,0.6) -- (2.1,1.2) -- (2.5,1.2);
        \draw[quantumPath] (2.5,1.2) -- (2.8,0.6) -- (3,0.6);

        \draw[quantumPath] (0,0.6) -- (0.4,0.6) -- (0.7,0) -- (1.1,0);
        \draw[classicalPath] (1.1,0) -- (1.4,1.2) -- (1.8,1.2) -- (2.1,0.6);
        \draw[quantumPath] (2.1,0.6) -- (2.5,0.6) -- (2.8,0) -- (3,0);

        \draw[quantumPath] (0,0) -- (0.4,0);
        \draw[quantumPath] (0.4,0) -- (0.7,0.6) -- (1.1,0.6);
        \draw[quantumPath] (1.1,0.6) -- (1.4,0) -- (2.5,0);
        \draw[classicalPath] (2.5,0) -- (2.8,1.2) -- (3,1.2);

        \foreach \x/\y in {0.55/1.9, 1.25/1.9, 1.95/1.3, 2.65/1.3, 1.25/0.7, 1.95/0.7, 2.65/0.7, 0.55/0.1, 1.25/0.1, 2.65/0.1}
            \draw[<->, >=Stealth, ultra thin, gray] (\x, \y) -- (\x, \y+0.4);
    \end{scope}

    \node[font=\small\bfseries] at (4.25, -0.7) {Time (Iterations) $\to$};

\end{tikzpicture}
    \caption{Comparison between PT (left) and QePT (right) which displays our hypothesis that not all Markov chains must be Quantum-enhanced to produce a quantum speed-up.}
    \label{fig:(Qe)PT}
\end{figure*}
 
Tempering in parallel is an elegant algorithmic solution to a computational bottleneck: slow thermalisation in low temperatures. This is identical to the motivation for QeMCMC. Naturally, a method that employs both techniques should be expected to sample well in low temperatures, and thus perform well in optimisation. While PT employs multiple Markov chains each exchanging state with some probability every $k/2$ steps,  QePT employs Quantum-enhanced Markov chains with the same ability to swap configurations. PT relies on many Markov chains, each aiding the other to explore. If each chain could be improved, then the entire system would thermalise more efficiently.

\begin{algorithm}
\caption{Quantum-enhanced Parallel Tempering}
\label{alg:qept}\begin{algorithmic}[0]
\State $T_{ladder} \gets \text{Temperature ladder}$
\State $\{s^{(1)}, s^{(2)}, \dots, s^{(M)}\} \gets$ $M$ independent initial spin configurations
\State $k \gets$ Swap interval
\State $\text{step} \gets 0$ \Comment{Timesteps}
\State $M_q \gets$ number of quantum chains
\While{not converged}
\State $\text{step} \gets \text{step} + 1$\For{$i = 1$ \textbf{to} $M$} \Comment{Parallel MCMC Step}

\If{$i\leq M_q$} 
\State $s^{(i)^\prime} \gets \text{QuantumProposal}(s^{(i)})$

\Else
\State $s^{(i)^\prime} \gets \text{LocalProposal}(s^{(i)})$
\EndIf
\State $\Delta f \gets f(s_i') - f(s_i)$

\State $A(s^{(i)^\prime} \mid s^{(i)}) \gets \min \left(1, \frac{\mu(s^{\prime})}{\mu(s)} \frac{Q(s^{(i)} \mid s^{(i)^\prime})}{Q(s^{(i)^\prime} \mid s^{(i)})}\right)$
\State $u \gets \text{Uniform}(0,1)$
\If{$u < A(s^{(i)^\prime} \mid s^{(i)})$}

\State $s^{(i)} \gets s^{(i)^\prime}$
\EndIf
\EndFor
\If{$\text{step} \pmod k == 0$} \Comment{Replica Exchange Step}
    \State $start \gets 2 - ((\text{step}/k) \pmod 2)$ 
    \For{$i = start, start+2, \dots$ while $i < M$}
        \State $j \gets i + 1$
        \State $\Delta \beta \gets (1/T_i - 1/T_j)$
        \State $\Delta f_{ij} \gets f(s^{(i)}) - f(s^{(j)})$
        \State $v \gets \text{Uniform}(0,1)$
        \If{$v < \exp(\Delta \beta \cdot \Delta f_{ij})$}
            \State \text{Swap } $s^{(i)} \leftrightarrow s^{(j)}$
        \EndIf
    \EndFor
\EndIf
\EndWhile
\State \Return $s$ \text{associated with } $T_{\min}$
\end{algorithmic}
\end{algorithm}

The spectral gap of a PT chain, $\delta_{PT}$, is considerably more complex than that of just one chain. Instead of tracking how a single chain traverses through $2^n$ configurations, one must consider $M$ replicas, meaning that transition at each step of the algorithm can go between $2^{mn}$ overall configurations. The spectral gap now no longer depends just on the $\delta$ of individual chains, but also of how well chains can communicate. In fact, there is classical literature which suggests that given chains restricted to local modes of the distribution, the overall chain will converge efficiently if there is sufficient overlap between them \cite{lee2023improved, woodard2009conditions}.

Of course, for many distributions, low-temperature chains still form a bottleneck as local minima result in drastically reduced overlap between poorly thermalised chains. This means that QePT is likely to result in a useful advantage over PT, when quantum-enhanced chains are employed only on the lowest chains. It is for this reason that we choose to study the convergence of our algorithm, where only a small subset of low-temperature chains are quantum-enhanced \footnote{Often PT is considered as replica Markov chains themselves swapping, however it is equivalent to think of them as instead swapping their configurations. We choose the latter convention.}. This hypothesis has large implications for the practicality of QePT in an HPC-QC environment.

\subsubsection{QePT Proof of principle}
\label{sec:QePT_pop}
As with the numerical results presented for QeSA (Sec. \ref{sec:QeSA_pop}), we take $T_\text{ladder}$ to be between $T_{\text{high}} = 10$ and $T_{\text{low}} = 0.1$, and of the form $T_i = T_{\text{high}} \times e^{-kx}$. We also take the number of steps between swap attempts to be $k=n$. In order to fairly compare methods, we must find the length of the PT chain that optimises the computational effort, Eq. \ref{eqn:effort}. To simplify the analysis we choose a very small system of $4$ replicas, however we note that for large instances many replicas are often employed. Although we only consider the case of $4$ rungs, the number of quantum-enhanced Markov chains ($M_{\text{q}}$) is varied, and the $M_{\text{q}}$ lowest temperature chains are taken to be quantum. The quantum hyper-parameters for QePT are taken to be the same as QeSA. The algorithm is implemented in \cite{Ferguson_Quantum-enhanced_Parallel_Tempering}.

\begin{figure}[h]
    \centering
    \includegraphics[width=1\linewidth]{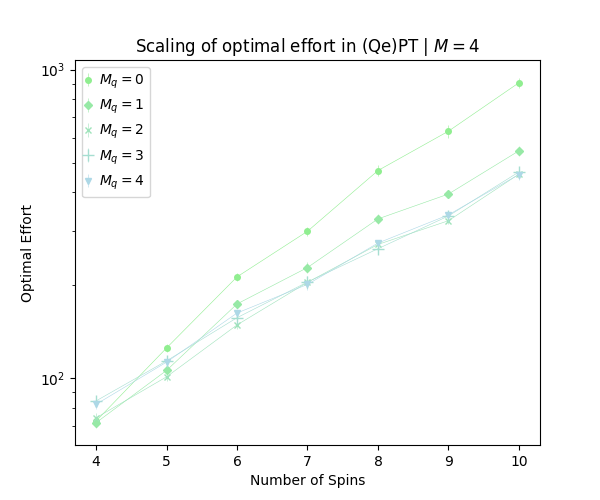}
    \caption{Scaling of optimal effort, for PT ($M_q = 0$, green, triangle) and QePT ($M_q = 4$, blue, circles). The hybrid cases with $M_q$ quantum-enhanced chains are also included, and represented by varying colours between green and blue. Optimal efforts are found using the same method as (Qe)SA, while 100 new Ising model instances are used in this graph. }
    \label{fig:QePT_scaling}
\end{figure}

Once appropriate annealing rates are found, the scaling of computational effort for QePT and PT can be compared in Fig. \ref{fig:QePT_scaling}. It is clear that not all chains need to be quantum to achieve an advantage over classical approaches. In fact, quantum-enhancement of only the lowest chains provides a tangible improvement. This ability to flexibly quantum-enhance only a small part of the overall computational load is an extremely exciting prospect for HPC-QC.

\section{Discussion}\label{sec:discussion}

Utilising Quantum-enhanced Markov chains in both SA and PT clearly has the potential for a quantum scaling advantage in combinatorial optimisation. However, quantifying the scaling of these heuristic approaches requires experimentation on real quantum hardware. The high sampling overhead of the one-shot-per-step protocol renders classical simulation of quantum-enhanced chains computationally prohibitive, even for intermediate system sizes. Conversely, however, it is well suited to practical quantum computation. 

In both SA and PT, we observed the hypothesised scaling advantage when using quantum-enhanced chains, however the parallelised nature of PT makes it the main algorithm of interest due to its potential in HPC-QC. In the example (Fig. \ref{fig:QePT_scaling}) of PT with $4$ replicas, using quantum-enhanced chains clearly reduces the effort required to find the global minima. However, when high temperature quantum-enhanced replicas are replaced by classical chains, much of this advantage remains. In fact, our results indicate a diminishing marginal return for quantum resources; specifically, configurations utilizing only $2$ quantum-enhanced replicas achieved a similar performance to the fully quantum-enhanced systems. This type of heterogeneous parallel tempering suggests a highly favourable resource-to-performance trade-off for HPC-QC.

There exists many practical questions in terms of running this algorithm. For example, how should one pick which replicas should be quantum-enhanced? A similar choice is made to decide how many replicas in any PT, and as with any heuristic, the answer is a combination of maximisation of available resources and trial-and-error. We know from previous literature \cite{layden2023, ferguson2025quantum} that Sherrington-Kirkpatrick models quickly become harder to thermalise below $t \approx 1$, and in many physical models, thermalisation is more difficult beyond a phase transition. In future, some knowledge of the problem formulation could thus be employed to suggest a critical temperatures, below which quantum chains become useful.

Alongside the traditional hyper-parameters, this family of quantum-enhanced methods require well-chosen $\gamma$ and $t$. Empirical findings demonstrate that identifying good parameters is possible in practise, while there are some analytical arguments to suggest that finding parameters that improve convergence may be difficult \cite{christmann2025quantum, orfi2024quantum}. In QePT, different replicas may benefit from different hyper parameters, just as optimal values vary for QeMCMC at different temperatures.

To further leverage classical compute, one could hybridize the chains themselves. In other words, a given Markov chain could, with some probability, perform either a classical or a quantum update proposal, reducing the reliance on the available quantum  resources, while still gaining some advantage from a chains decreased locality. In many cases, a combination of two ergodic proposal mechanisms can be combined, as is often used classically to balance local and more global proposals \cite{tierney1994markov}. This is just one practical consideration, among many, that any user will have to examine when implementing QePT, however current HPC-QC is not at a stage to realize these techniques.

Quantum devices that are available now, and in the near future, have some undesirable qualities which necessitate careful consideration. First, is the presence of quantum errors, however it should be stressed that QeMCMC is expected to be naturally resilient. In the limit of fully depolarising noise, QeMCMC proposals uniformly sample configurations, which is still a valid MCMC proposal. Another issue is the slow runtime of quantum processors. Although the original proposal of QeMCMC empirically argues for reduced  thermalisation over classical MCMC, the quantum device takes longer to execute a circuit than a classical computer takes to flip a bit. This initial disadvantage may not be overcome by the advantages of fast thermalisation until one considers very large systems, however by offloading compute (such as high temperature replicas) to classical hardware, this is mitigated.

Future work should carefully compare, our methods against other heuristics, both quantum and classical. For example, even within the MCMC family, different proposal mechanisms exist which we have not considered here \cite{wolff1989collective, park2017rapid, swendsen1987nonuniversal}. Comparison with established quantum approaches, such as VQAs or QA, should also be carried out, however directly comparing the overheads (both quantum and classical) of vastly different algorithms is rarely trivial. Combinatorial optimisation encompasses a vast array of computational problems, many of which possess specific structural properties that admit highly efficient classical algorithms. Consequently, while quantum-enhanced methods may prove powerful for glassy landscapes, they may not be the optimal choice for all problem classes. It is therefore critical to evaluate the performance of our methods across a diverse selection of problem types and instances to define their regime of utility.

\section{Conclusion}

This work proposes two novel methods for combinatorial optimisation: Quantum-enhanced Simulated Annealing (QeSA) and Quantum-enhanced Parallel Tempering (QePT). Both are quantum extensions of classical techniques, which in turn rely on Markov Chain Monte Carlo subroutines. Inspired by \cite{layden2023}, we enhance these algorithms by employing quantum real-time evolution to propose new configurations for Markov chains to explore. We perform proof of principle numerics on simulated hardware, from which it is clear that our quantum-enhanced approaches both show considerable potential. In particular, QePT has inherent utility as a heterogeneous approach to parallel tempering by employing both classical and quantum chains in parallel. 

While this study focuses on simulated annealing and parallel tempering, related classical algorithms such as population annealing and simulated tempering may offer analogous pathways to quantum enhancement. There is also potential to investigate other extensions of the quantum-enhanced literature such as coarse graining, constrained systems, quantum annealing and quantum inspired approaches \cite{arai2025quantum, christmann2025quantum, ferguson2025quantum, ferguson2025dynamics}. Current quantum hardware seldom supports the classical integration that is required to test our algorithms and classical simulation is restricted in what it can teach us. As hardware architectures evolve tighter quantum-classical integration, the practical utility of this family of methods will become increasingly verifiable.

\section*{Acknowledgments}
We would like to thank Feroz Hassan for useful discussions.
S.F. acknowledges funding from EPSRC DTP studentship grant EP/W524311/1 and P.W. funding from EPSRC grants  EP/X026167/1 and EP/Z53318X/1.

\bibliography{bib.bib}

\appendix

\onecolumngrid

\section{Extended numerics}\label{app:SA_numerics}

\begin{figure*}[!h]
    \centering
    \includegraphics[width=0.8\linewidth]{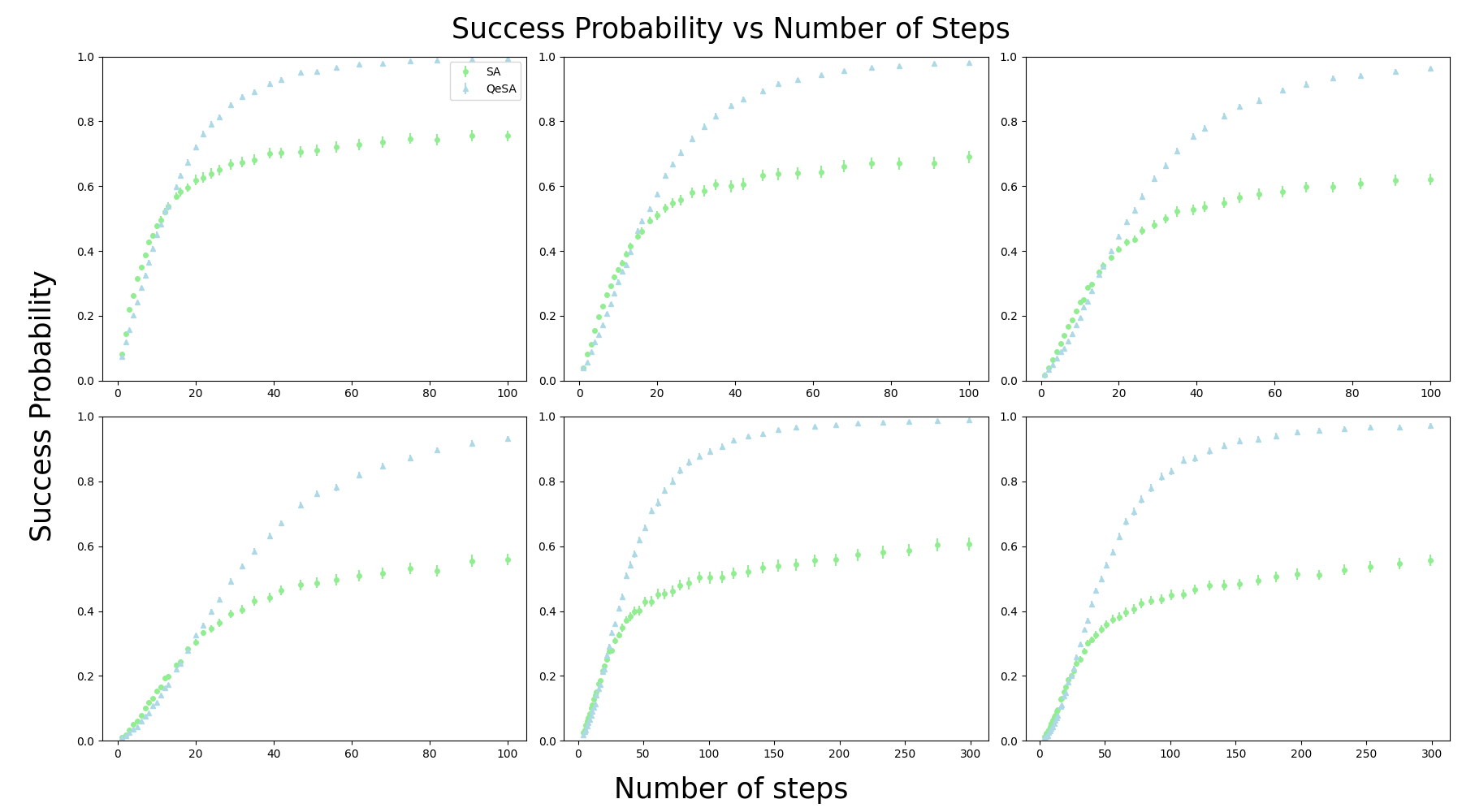}
    \caption{Probability of finding global minim in a single SA (green, circle) or QeSA (blue, triangle) run against the number of steps of the run. Naturally, slower anneals are more likely to find the optimal configuration, but at the cost of extra steps. Each subplot here is equivalent to Fig. \ref{fig:10_hopsprobs}, but for $n = \{4,5,6,7,8,9\}$. Note that x scale changes between sub-graphs. }
    \label{fig:SA_ps_all}
\end{figure*}

\begin{figure*}[!h]
    \centering
    \includegraphics[width=0.8\linewidth]{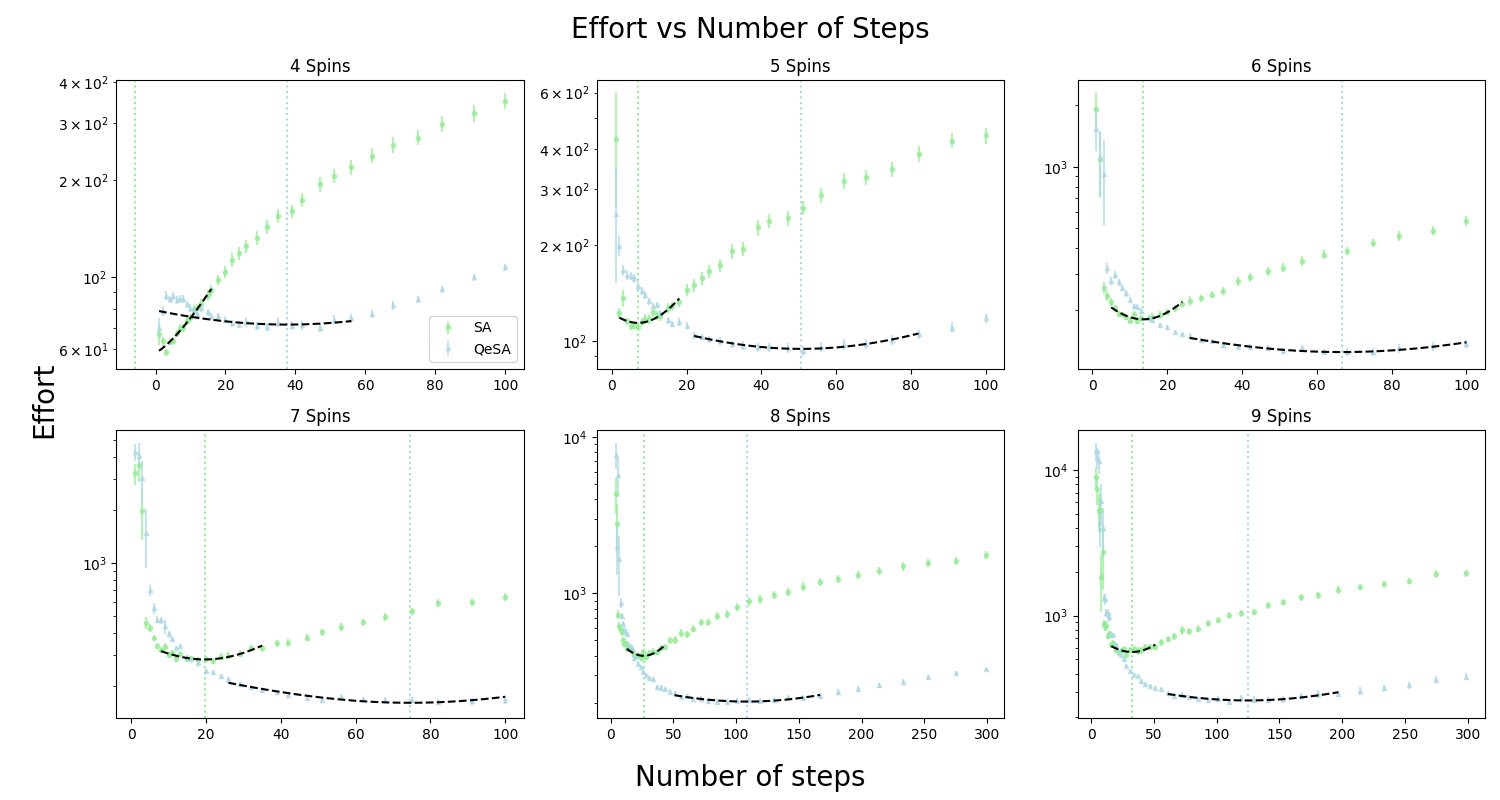}
    \caption{Computational effort required to find the global minim in a single SA (green, circle) or QeSA (blue, triangle) run against the number of steps of the run. There is an optimal length of an annealing run, after which it is more efficient to run extra short anneals, rather than fewer longer runs. Each subplot here is equivalent to Fig. \ref{fig:10_effort}, but with $n = \{4,5,6,7,8,9\}$. For $n=4$, the triviality of the system means that extremely short anneals are useful, and the quadratic fit is less effective. Note that both x and y scales change between sub-graphs.}
    \label{fig:SA_effort_all}
\end{figure*}

\begin{figure*}[h]
    \centering
    \includegraphics[width=0.5\linewidth]{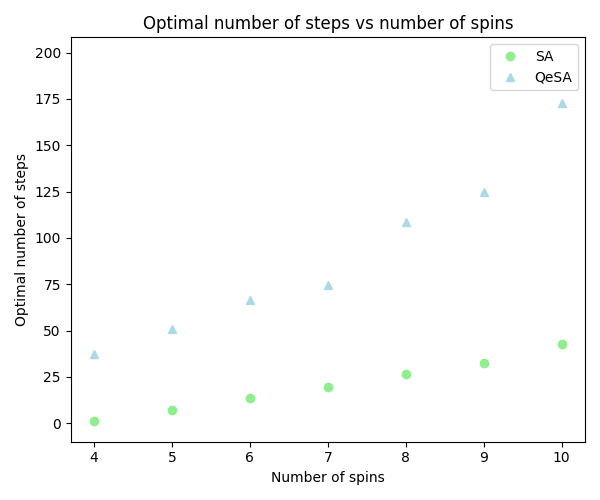}
    \caption{Optimal number of steps (found by quadratic fits in Fig. \ref{fig:10_effort} and Fig. \ref{fig:SA_effort_all}) against $n$. }
    \label{fig:SA_nhops}
\end{figure*}

In this section, we provide numerical results for (Qe)SA that were not included in the main text. In Fig. \ref{fig:SA_ps_all}, the probability of success is plotted against number of steps for $\{5,6,7,8,9\}$ spins. In Fig. \ref{fig:SA_effort_all}, the resulting effort for each number of spins is plotted. Finally, in Fig. \ref{fig:SA_nhops}, the optimal number of steps (found by quadratic fit in Fig. \ref{fig:SA_effort_all}) is plotted for each $n$ that we consider.

We also include the optimal number of steps vs $n$ for PT, however we do not include the probability and effort plots, as the volume of data is too large. The method used in the case of PT is identical to that of SA.

\begin{figure*}[h]
    \centering
    \includegraphics[width=0.5\linewidth]{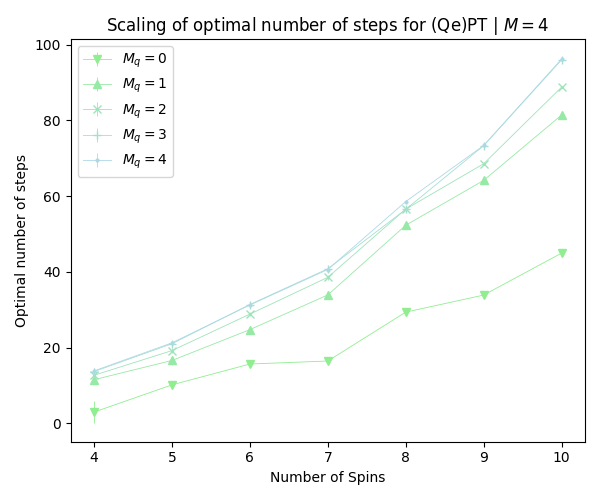}
    \caption{Optimal number of steps (found by quadratic fits similar to Fig. \ref{fig:10_effort}, but for PT) against $n$. }
    \label{fig:PT_nhops}
\end{figure*}

\end{document}